# Impact of Covid-19 Pandemic on Water Pollution in Indian Rivers-A Case Study

Amardeepak Mahadikar[1*], Krishna Anand[2], Chandra S. Reddy[3]

[1] Department of Mechanical Engineering, Cambridge Institute of Technology - North Campus, Bangalore 562110, India
[2] Department of Artificial Intelligence, Anurag University, Hyderabad 500088, India
[3] Department of Mathematics, Cambridge Institute of Technology -North Campus, Bangalore 562110, India

Corresponding Author Email: amardeepakm1@gmail.com

**ABSTRACT**

Some of the important critical parameters for assessing the water quality like pH (Hydrogen ion concentration), DO (Dissolved Oxygen), BOD (Biological Oxygen Demand), etc., were monitored at different locations in some major Indian rivers. The results obtained from the study reveals that the critical parameters had increasing values in some monitoring locations, decreasing values, and no variation in values at some other places. It is recommended to have a pH value above 7, higher values of DO, lower values of BOD & FCC (Faecal Coliform Content) for improved water quality. Overall, the water quality improved in most Indian rivers. There was no discharge of industrial wastes, hotels/restaurants wastes, immersing of idols during religious festivals, etc., to the rivers during the COVID-19 lockdown. Therefore, enforcement of strict regulations by the Government of India for disposal of wastes produced from industrial & domestic activities can significantly reduce the water pollution levels in the Indian rivers.

## 1. INTRODUCTION

COVID-19 stands for Corona Virus Infectious Disease, whose year of occurrence is 2019. It is caused by the pathogen Severe Acute Respiratory Syndrome Corona Virus-2 (SARS-COV-2) belonging to the β-subgroup of the Corona virus family. The disease was first diagnosed in Wuhan city, Hubei province of China, which later spread its tentacles to over 220 countries and territories around the world. The Government of India imposed a nationwide lockdown since midnight of 24th March to restrict the spread of the deadly Corona virus disease Covid-19. The World Health Organization (WHO) declared it a global pandemic of international concern on 30th Jan 2020. It is found that human-to-human transmission is mainly by close contact with an infected person through coughing, sneezing, respiratory droplets. However, there are cases reported of transmission by viral shedding via faeces [1, 2]. Some of the common symptoms of COVID-19 infection are fever, headache, fatigue, dry cough, respiratory distress, vomiting, diarrhea, etc.

Water pollution is a major global problem that gives rise to water-borne diseases such as cholera, typhoid, hepatitis, etc., [3, 4]. As per estimates, every year globally, around 1.7 million children below five years of age die, and 38 million Indians suffer from various water-borne diseases. Before the COVID-19 lockdown, major Indian rivers and lakes were heavily polluted due to human activities and getting difficult to be treated [5, 6]. As reported by the Central Pollution Control Board (CPCB) of India, 40 million litres of wastewater enter the rivers and other water bodies every day. Only 37% is treated adequately [7]. The rapid urbanization has caused contamination of 70% of freshwater sources in India making them unfit for consumption.

The imposition of lockdown to contain the spread of the virus resulted in restrictions on public transportation, commercial and industrial activities that positively impacted the environment and were a blessing in disguise to Mother Nature [8-10]. Improvement in water quality across the river water environment due to reduced economic activities resulted in less pollutants discharged to the rivers [11-13]. Ganga river water quality has shown significant improvement during the COVID-19 lockdown and was found suitable for bathing at most monitoring stations. The enforcement of nationwide lockdown also improved the health of other major Indian rivers.

In this paper authors focused to study the effect of COVID-19 and its consequences on geographical conditions. It focuses on to study water pollution in the Indian rivers and the characteristics changes. Section 2 discusses the methodology implemented to study characteristic behaviors of water in the rivers. A simple statistical model is presented to measure the parameters. Section 3 deals the results and analysis of the work presented in this paper. Results are presented in graphical and tabular forms, before and after of the COVID period. It shows that during this period natural conditions are much improved. Section 4 presents the conclusion part of the paper.

## 2. MATERIAL AND METHODS

For the N independent samples, $X_n$, where n = 1, 2, …, N represents independently distributed random variables tested in the experiments. Then, the mean value for this experiment can be expressed as an estimation of the average of all samples [14-18].

i.e., $\hat{X} = \frac{1}{N}\sum_{n=1}^{N} X_n$

This represents the effect of estimation of the mean sample [19-21].

The estimation in terms of variance is expressed as:

$$E[\hat{X}_N] = E\left[\frac{1}{N}\sum_{n=1}^{N} X_N\right]$$
$$= \frac{1}{N}\sum_{n=1}^{N}[X_N] = \hat{X}$$

Since the system is unbiased, hence the measurement function for the mean estimator is further expressed in variance form as:

$$E\left[\left(\hat{X}_N - \hat{X}\right)^2\right] = E\left[\hat{X}_N^2 - 2\hat{X}\hat{X}_N + \hat{X}^2\right]$$
$$= -\hat{X}^2 + E\left[\frac{1}{N}\sum_{n=1}^{N} X_n \frac{1}{N}\sum_{m=1}^{N} X_m\right]$$
$$\sigma_{X_N}^2 = -\hat{X}^2 + \frac{1}{N^2}\sum_{n=1}^{N}\sum_{m=1}^{N} E[X_n X_m]$$

## 3. RESULT AND DISCUSSION

### 3.1 River Ganga

The imposition of lockdown to control the spread of the Corona virus has resulted in a significant reduction in water pollution levels of the Ganga River by 25 to 30%. The biggest and highly contaminated river runs in northern parts if India. On average, the DO concentration increased by 20 to 30%, and BOD decreased by 35 to 40%, based on the studies conducted to monitor the water pollution levels of the sacred river. The quantity of decaying organic matter indicates the BOD level. A lower concentration of BOD and a higher level of DO indicate good water quality. A reduced level of DO in water severely affects aquatic life. The prescribed standard for the survival of aquatic life is BOD less than 3 mg/litre and DO above 5 mg/litre.

The measured DO value of river Ganga on 6th March 2020 at Varanasi was 3.8 mg/litre, and on 4th April 2020, 6.8 mg/litre indicating an improvement of 79%. Table 1 shows the data collected by Central Pollution Control Board (CPCB) on 28th March 2020, indicating satisfactory levels of recorded values of DO, BOD & pH at Upstream & Downstream Ganga.

**Table 1.** CPCB data of measured values of pH, DO & BOD

| Monitoring Station | Parameter | Measured values |
|---|---|---|
| Ganga river Upstream | pH | 7.90 |
|  | DO | 8 mg/litre |
|  | BOD | 2.1 mg/litre |
| Ganga river Downstream | pH | 7.91 |
|  | DO | 7.90 mg/litre |
|  | BOD | 1.21 mg/litre |

The observations of river Ganga in Uttar Pradesh (UP) showed satisfactory results. Totally, 14 locations were monitored. Increasing values of DO and BOD were observed in 8 and 4 locations. Decreasing DO, BOD & FCC values were observed in 6, 9 & 10 locations, respectively. No change in BOD value was observed in 1 location. The average values of critical parameters monitored in 14 locations during pre-lockdown and lockdown are shown in Table 2 below. Graphical results are presented in Figure 1.

**Table 2.** Average values of critical parameters at river Ganga monitored in UP

| Parameter | Pre-lockdown | Lockdown |
|---|---|---|
| pH | 5.95 | 8.05 |
| DO (mg/litre) | 9.3 | 9.4 |
| BOD (mg/litre) | 2.8 | 2.45 |

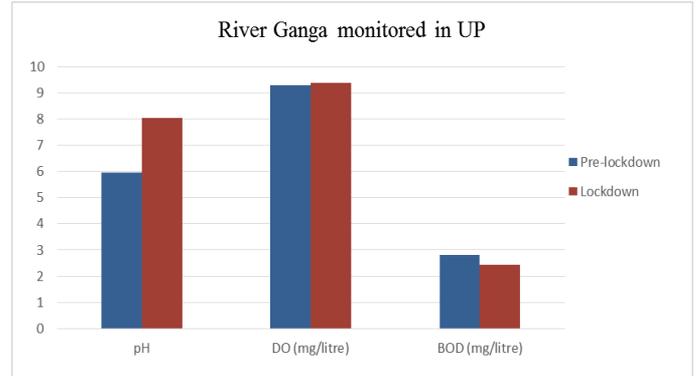

**Figure 1.** pH, DO & BOD values of Ganga river monitored in UP during pre-lockdown & lockdown

### 3.2 River Yamuna

In Himachal Pradesh (HP), the river Yamuna was monitored at 4 locations. Increasing values of DO were observed in 4 locations, decreasing values of BOD & FCC in 4 & 3 locations. No change in FCC was observed in 1 location. The average values of critical parameters measured in 4 locations are shown in Table 3. These results are presented in the form of bar charts in Figure 2.

**Table 3.** Average values of critical parameters at river Yamuna monitored in HP

| Parameter | Pre-lockdown | Lockdown |
|---|---|---|
| pH | 7.05 | 7.2 |
| DO (mg/litre) | 8.75 | 8.95 |
| BOD (mg/litre) | 0.7 | 0.4 |

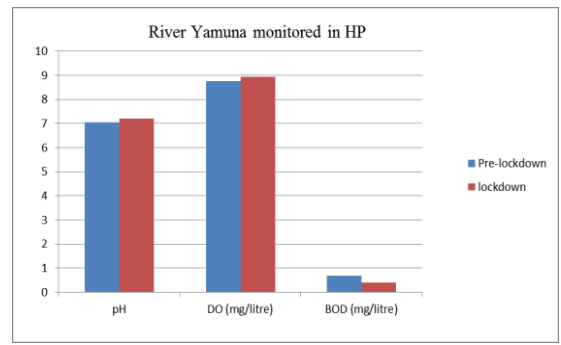

**Figure 2.** pH, DO & BOD values of river Yamuna monitored in HP during pre-lockdown & lockdown

### 3.3 River Brahmaputra

In Assam, the river Brahmaputra was monitored for water quality at 8 locations during pre-lockdown and 10 locations during the lockdown. Increasing values of DO, BOD & FCC were observed in 3, 1, 4 locations and decreasing values in 5, 7, 2 locations, respectively. Figure 3 represents graphical form of the results for the river Brahmaputra. No change in FCC was observed in 1 location. The average values of critical parameters recorded in Assam are shown in Table 4.

**Table 4.** Average values of critical parameters at river Brahmaputra monitored in Assam

| Parameter | Pre-lockdown | Lockdown |
|---|---|---|
| pH | 7.85 | 7.65 |
| DO (mg/litre) | 7.75 | 8.45 |
| BOD (mg/litre) | 2 | 1.6 |

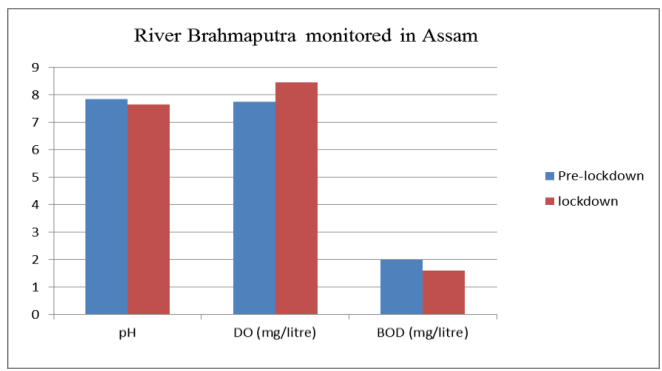

**Figure 3.** pH, DO & BOD values of river Brahmaputra monitored in Assam during pre-lockdown & lockdown

### 3.4 River Godavari

The river Godavari flows in the sadharan parts of India was monitored at 14 locations in Maharashtra and Andhra Pradesh to check for water quality during pre-lock down and lockdown. It was observed there were increasing values of DO, BOD & FCC in 9, 3 & 1 locations and decreasing values in 3, 10 & 4 locations, respectively. The values of DO, BOD & FCC remained unchanged in 2, 1 & 9 locations. Table 5 below shows the average values of critical parameters measured in 14 locations during the pre-lockdown & lockdown periods. These results are presented in Figure 4 for Godavari river, one of the key rivers in south India.

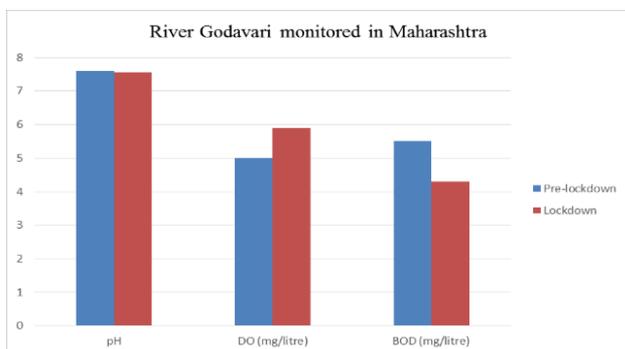

**Figure 4.** pH, DO & BOD values of river Godavari monitored in Maharashtra during pre-lockdown & lockdown

**Table 5.** Average values of critical parameters at river Godavari monitored in Maharashtra

| Parameter | Pre-lockdown | Lockdown |
|---|---|---|
| pH | 7.6 | 7.55 |
| DO (mg/litre) | 5 | 5.9 |
| BOD (mg/litre) | 5.5 | 4.3 |

### 3.5 River Narmada

The river Narmada was monitored for water quality in Gujarat at 5 locations. Increasing values of DO and FCC were observed at 3 & 2 locations. Decreasing DO, BOD & FCC values were observed in 2, 2 & 3 locations, respectively. No change in BOD value was observed at 3 locations. The average values of critical parameters measured in 5 locations during the pre-lockdown & lockdown period are shown in Table 6. These results are presented in graphical form in Figure 5.

**Table 6.** Average values of critical parameters at river Narmada monitored in Gujarat

| Parameter | Pre-lockdown | Lockdown |
|---|---|---|
| pH | 7.85 | 7.45 |
| DO (mg/litre) | 7.5 | 7.65 |
| BOD (mg/litre) | 0.7 | 0.6 |

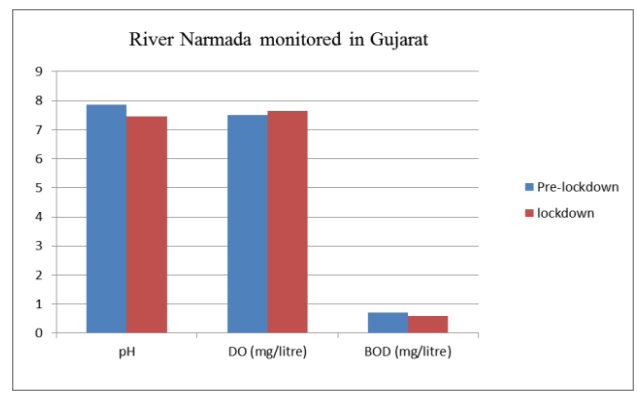

**Figure 5.** pH, DO & BOD values of river Narmada monitored in Gujarat during pre-lockdown & lockdown

### 3.6 River Krishna

In Andhra Pradesh (AP), the water quality of the river Krishna was monitored at 8 locations in Karnataka and Andhra Pradesh. It passes through different locations in the south India in various states. The DO, BOD & FCC showed increasing values in 4, 1 & 2 locations, respectively. Decreasing values of DO were observed in 3 locations and BOD in 4 locations. No variation in DO values was observed in 1 location, BOD in 3 locations & FCC in 6 locations. The average values of critical parameters measured in 8 locations are shown in Table 7 and graphical results are presented in Figure 6.

**Table 7.** Average values of critical parameters at river Narmada monitored in Gujarat

| Parameter | Pre-lockdown | Lockdown |
|---|---|---|
| pH | 7.4 | 7.55 |
| DO (mg/litre) | 6 | 6 |
| BOD (mg/litre) | 1.8 | 1.6 |

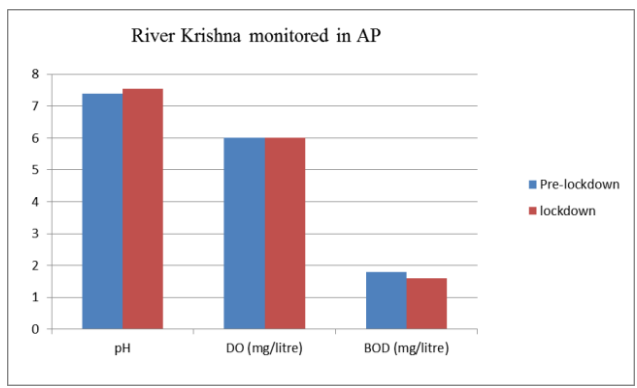

**Figure 6.** pH, DO & BOD values of river Krishna monitored in AP during pre-lockdown & lockdown

### 3.7 River Cauvery

The water quality of the river Cauvery was monitored at 22 locations in Karnataka. Increasing values of DO were observed in 21 locations, decreasing values of BOD & FCC in 20 & 21 locations, respectively. No variation in DO value was observed in 1 location, BOD values in 2 locations, and FCC value in 1 location. Table 8 shows the average values of critical parameters measured in 22 locations during the pre-lockdown & lockdown periods. It is one of the key rivers in Karnataka and Tamil Nadu states of India. These results are presented in graphical form in Figure 7.

**Table 8.** Average values of critical parameters at river Cauvery monitored in Karnataka

| Parameter | Pre-lockdown | Lockdown |
|---|---|---|
| pH | 7.6 | 8 |
| DO (mg/litre) | 6.65 | 7.2 |
| BOD (mg/litre) | 2.05 | 1.45 |

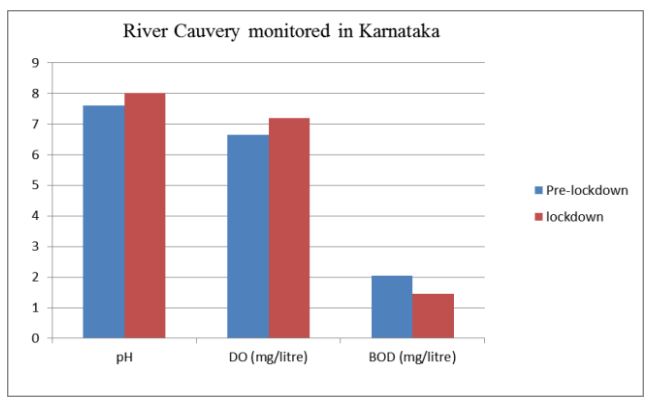

**Figure 7.** pH, DO & BOD values of river Cauvery monitored in Karnataka during pre-lockdown & lockdown

### 4. CONCLUSIONS

In this manuscript, the authors attempted to address the impact of COVID-19 on water pollution in Indian rivers. The critical parameters for assessing the water quality like pH, DO & BOD in some significant Indian rivers during pre-lockdown and lockdown are studied in this work. The pollution levels reduced in the major Indian rivers due to the Covid-19 lockdown, thereby showing a remarkable improvement in water quality due to a complete halt to tourism, pilgrimage, and industrial activities. However, domestic sewage contributed to the pollution of the water bodies.

The findings suggest that the critical parameters monitored during the lockdown period showed satisfactory levels. These changes may be temporary because, after the lockdown, industrial and human activities will increase, due to which more pollutants will be discharged to the water bodies. Therefore, it is necessary for governments in general/ individuals in particular, to learn from the environmental impact due to lockdown and adopt proper measures to reduce pollution on a long-term basis for the welfare of society.